\documentclass[apsrev,prb,twocolumn,showpacs,preprintnumbers,amsmath,amssymb,groupaddress]{revtex4}

\usepackage{graphicx}
\usepackage{amsmath}
\usepackage{amsfonts}
\usepackage{color}

\begin{document}

\title{Non-linear resistivity and heat dissipation in monolayer graphene}

\author{A. S. Price, S. M. Hornett, A. V. Shytov, E. Hendry, D. W. Horsell}
\affiliation{School of Physics, University of Exeter, Exeter EX4 4QL, UK}

\begin{abstract}

We have experimentally studied the nonlinear nature of electrical conduction in monolayer graphene devices on silica substrates. This nonlinearity manifests itself as a nonmonotonic dependence of the differential resistance on applied DC voltage bias across the sample. At temperatures below $\sim$70K, the differential resistance exhibits a peak near zero bias that can be attributed to self-heating of the charge carriers. We show that the shape of this peak arises from a combination of different energy dissipation mechanisms of the carriers. The energy dissipation at higher carrier temperatures depends critically on the length of the sample. For samples longer than~10$\mu$m the heat loss is shown to be determined by optical phonons at the silica-graphene interface.

\end{abstract}
\maketitle

Graphene is a promising material for electronic applications because of its high electrical and thermal conductivities \cite{GeimScience09}. A limiting intrinsic room-temperature mobility of graphene field-effect transistors of $\sim$$2 \times 10^5$ cm$^2$/Vs has been estimated \cite{ChenNatNano08} and approached experimentally \cite{BolotinPRL08}. Its high thermal conductivity\cite{BalandinNanoLett08}, far exceeding that of copper, holds great promise for heat management in electronics \cite{PopNanoRes2010}. Fully realizing this potential requires an understanding of the scattering mechanisms in graphene, particularly electron-phonon scattering, which becomes increasingly important with increasing temperature and device currents.

Several different techniques have been used to study the coupling between charge carriers and phonons in graphene. Measurements of the temperature dependence of the resistance have shown that at low temperatures, $T$$<$ 150 K, electron-phonon scattering is dominated by longitudinal acoustic phonons \cite{StauberPRB07,HwangPRB08,MorozovPRL08,ChenNatNano08}, resulting in a linear temperature dependence of the resistivity. At higher temperatures the resistivity becomes strongly temperature dependent, possibly due to ripples in the graphene membrane \cite{MorozovPRL08} or coupling with remote interfacial phonons (RIPs) at the SiO$_2$ surface \cite{ChenNatNano08,FratiniPRB08}. This latter scattering mechanism has been put forward as an important heat dissipation mechanism in carbon nanotube and graphene devices \cite{RotkinNanoLett09,FreitagNanoLett09,ViljasPRB11}. Measurements of the current saturation in graphene lend support to the hypothesis that these RIPs are an important scattering mechanism \cite{MericNatNano08}, although another study identifies scattering with optical phonons intrinsic to the graphene lattice \cite{BarreiroPRL09} as the dominant saturation mechanism.

In this paper, we report electrical measurements of monolayer graphene field-effect transistors in the high-current regime at temperatures between 4 and 200 K. We show that overheating of the charge carriers results in a nonmonotonic dependence of the differential resistance on source-drain bias. By correlating this dependence with the measured temperature dependence of the differential resistance, we test different models of thermal transport and electron-phonon coupling, and demonstrate that energy dissipation is dominated by coupling to the remote interfacial phonons in the silica substrate.

\begin{figure}[h]
\begin{center}\leavevmode
\includegraphics[width=0.8\linewidth]{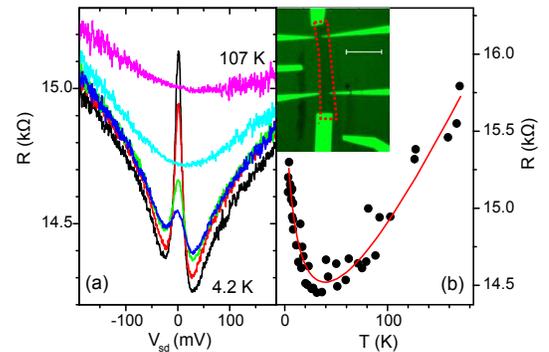}
\caption{(color online). (a) Differential resistance as a function of DC source-drain bias measured at different bath temperatures: $T_B$ = 4.2, 17, 26, 35, 83, and 107 K from bottom to top. (b) Dependence of the zero-bias differential resistance on the bath temperature. All measurements are taken at the Dirac point. The solid red line is a guide to the eye. Inset: optical micrograph of sample A, with the flake indicated by a dotted outline; the scale bar is 10 $\mu$m.}
\label{PaperFig1}
\end{center}
\end{figure}

Graphene flakes were mechanically exfoliated onto conductive silicon substrates covered by a 300 nm thick SiO$_2$ layer, allowing carrier density control by application of a voltage between flake and substrate. Flakes were identified as single layers by optical contrast measurements and Raman spectroscopy, and were electrically contacted with Cr/Au (5/50 nm) to allow four-terminal measurement of the differential resistance. This resistance, $dV/dI$, was measured by a low-frequency ($\sim$10 Hz) lock-in technique using sufficiently small AC current to prevent unintentional self-heating (typically less than 5 nA, corresponding to an AC source-drain voltage $\sim$100 $\mu$V). The differential resistance was measured as a function of DC voltage, $V_{sd}$, applied between source and drain contacts. We consider below the results of two representative samples of different lengths, which demonstrate different thermal transport regimes. Sample A is shown in the inset of Fig.~\ref{PaperFig1}(b), and is 3.7 $\mathrm{\mu m}$ wide and 16 $\mathrm{\mu m}$ between voltage probes, with a typical mobility of 5700 $\mathrm{cm^2/Vs}$. Sample B is a shorter sample, 4 $\mu$m wide and 10 $\mu$m in length, with a lower mobility of 2500 cm$^2$/Vs.

Figure \ref{PaperFig1}(a) shows the differential resistance of sample A as a function of DC source-drain bias for different bath temperatures. These measurements were taken in an atmosphere of dry helium, with the gate voltage, $V_G$, tuned to the Dirac point. At the highest temperature shown, $T_B$ = 107 K, a small increase in $R$ is seen with increasing $V_{sd}$. As the bath temperature is decreased, we observe two features: the differential resistance becomes strongly dependent on $V_{sd}$, and a peak in the differential resistance is developed at zero bias. At low-temperatures the differential resistance behaves in a similar way to the temperature dependence of the zero-bias resistance $R_0(T)$, which is shown for the Dirac point in Fig.~\ref{PaperFig1}(b). At low temperatures, $R_0(T)$ exhibits a logarithmic behavior, due to the well known weak-localization and electron-electron interaction contributions \cite{TikhonenkoPRL08,KozikovPRB10}. The size of the low-temperature peak in $R(V_{sd})$ is $\sim$ $1 \mathrm{k\Omega}$, which is similar to the size of the low-temperature peak in $R_0(T)$. Furthermore, the peak in $R(V_{sd})$ occurs only for bath temperatures below the minimum in $R(T)$, which occurs at $T_B \approx 40$ K in Fig.~\ref{PaperFig1}(b): it is barely resolved in Fig.~\ref{PaperFig1}(a) at $T_B$ = 35 K, and is completely absent at $T_B$ = 83 K. These observations suggest that the nonmonotonic differential resistance originates from a hot-carrier effect.

\begin{figure}[h]
\begin{center}\leavevmode
\includegraphics[width=0.8\linewidth]{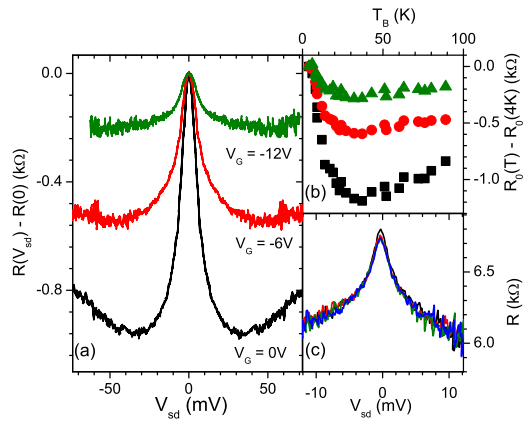}
\caption{(color online). (a) Differential resistance as a function of DC source-drain bias measured at $T_B$ = 4.2K, for different gate voltages, $V_G$ = 0, -6, -12V with respect to the Dirac point (bottom to top). (b) The temperature dependence of the zero-bias resistance, for the gate voltages shown in (a): $V_G$ = 0, -6, -12V, bottom to top. In (a) the zero-bias resistance has been subtracted from the curves to aid comparison, whist in (b) $R_0(T = 4.2\,\mathrm{K})$ has been subtracted. (c) Differential resistance as a function of source-drain bias, measured with different lengths between source and drain contacts, taken at 4.2 K. Measurements were taken for sample A in (a) and (b), and sample B in (c).}
\label{ndepRVsd}
\end{center}
\end{figure}

Further support for this hypothesis is provided by analyzing the evolution of the peak with carrier density. Figure \ref{ndepRVsd}(a) shows the change in differential resistance $R(V_{sd}) - R(0)$ against applied source-drain bias for different carrier densities. (A small asymmetric component of the differential resistance resulting from the DC source-drain bias modulating the gate voltage has been removed by symmetrization \cite{SOM}.) It can be seen that the peak in $R(V_{sd})$ is most pronounced at low carrier densities. This density-dependence of the peak size is also seen in the temperature dependence of the resistance below 50 K, Fig.~\ref{ndepRVsd}(b), suggesting this is also a manifestation of $R_0(T)$ due to a hot-carrier effect in $R(V_{sd})$. The features in Figs. \ref{PaperFig1} and \ref{ndepRVsd} have been seen in all monolayer graphene samples that we have measured.

To make a quantitative check of the hot-carrier hypothesis, one needs to analyze the production and distribution of Joule heat in the sample. In a micron-sized conductor this is determined by several characteristic lengths: the elastic impurity scattering length~$l$, electron-electron collision length $l_{e\text{-}e}$, and electron-phonon relaxation length~$l_{e\text{-}ph}$. At low temperatures, inelastic processes are suppressed by Pauli blocking, and the relaxation rates form a hierarchy, $l \ll l_{e\text{-}e} \ll l_{e\text{-}ph}$ ($l \sim$ 10 nm). Estimating the electron-electron collision time in sample A as $\tau_{e\text{-}e} \sim \hbar E_F / (k_B T)^2 \sim 0.2{\rm ns}$ at $V_G$ = 3V, one finds $l_{e\text{-}e} \sim \sqrt{D \tau_{e\text{-}e}} \sim 1{\rm \mu m}$. The electron-phonon relaxation length, which will be discussed later, can exceed 10 $\mu$m. In short conductors, $L \ll l_{e\text{-}e}$, electrons are scattered elastically, and no Joule heat is produced inside the conductor. When the conductor size exceeds several microns, so that~$l_{e\text{-}e} < L < l_{e\text{-}ph}$, the energy of electrons is thermalized by electron-electron collisions, and Joule heat is transferred only to the electrons. This heat is then dissipated via diffusion of electrons into the metal contacts, and the resulting electronic temperature distribution is described by a universal temperature profile \cite{RudinPRB95}:

\begin{equation}
T_e^2(x) = T_B^2 + \frac{3}{4 \pi^2}(eV_{sd})^2\left(1 - \frac{4x^2}{L^2}\right).
\label{EqRKTx}
\end{equation}

\noindent
Here $L$ is the sample length, and $x$ is the spatial co-ordinate running along the length of the sample from $-L/2$ to $L/2$. In longer samples, $L > l_{e\text{-}ph}$, electron and lattice temperatures are equilibrated via phonon emission.  Note that in graphene the phonon contribution to the specific heat, $C_{ph}\propto T^2$, dominates over the electron contribution, $C_e \propto T$, at all relevant temperatures. Therefore, nearly all Joule heat is transferred to the lattice by electron-phonon relaxation.

\begin{figure}[h]
\begin{center}\leavevmode
\includegraphics[width=0.9\linewidth]{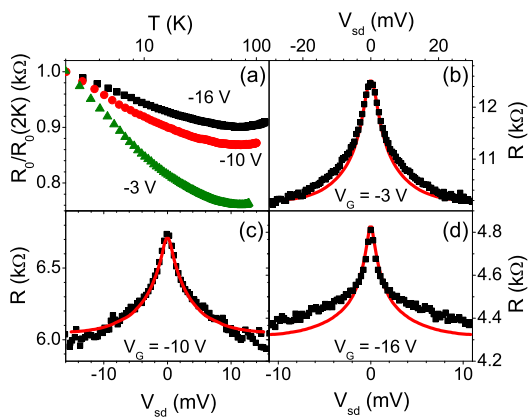}
\caption{(a) Temperature dependence of the normalized zero-bias resistance of sample B at $V_G$= -3, -10, -16 V with respect to the Dirac point. (b)--(d) differential resistance as a function of source-drain bias at different gate voltages, $V_G$ = -3, -10, -16 V, respectively. The bath temperature $T_B$ = 4 K in (b) and (c), and 2 K in (d). The symbols are measured values, and the red lines are fits using Eq.~(\ref{EqRKTx}).}
\label{IVfits}
\end{center}
\end{figure}

To test the Joule heating picture, we first analyze the data for the shorter 10 $\mu$m sample. Its zero-bias resistance is shown in Fig.~\ref{IVfits}(a) as a function of bath temperature. Neglecting electron-phonon relaxation we use the temperature profile in Eq.~(\ref{EqRKTx}) to compute the differential resistance:

\begin{equation}
R(V_{sd}) = \frac{1}{L}\int R(T_e(x))dx.
\label{EqRVsdTxIntegral}
\end{equation}

\noindent In Fig.~\ref{IVfits}(b)--(d) this model is compared with the differential resistance at three different gate voltages, $V_G$ = -3, -6, -12 V, with respect to the Dirac point. It can be seen that Eqs. (\ref{EqRKTx}) and (\ref{EqRVsdTxIntegral}) describe well the central peak in the differential resistance, which confirms our hypothesis about its origin and demonstrates the inefficiency of electron-phonon relaxation in the 10 $\mu$m sample. In taking $L$ in Eqs. (\ref{EqRKTx}) and (\ref{EqRVsdTxIntegral}) to be the distance between the voltage probes, rather than the distance between the source and drain contacts, we assume that carriers are held at the bath temperature at each metallic contact. In order to test this assumption, we measured $R(V_{sd})$ between the same voltage contacts in sample B, whilst using different contacts as source and drain such that the distance between the source and drain was varied between 25 and 50 $\mu$m, Fig.~\ref{ndepRVsd}(c). No effect is seen in $R(V_{sd})$ when changing the source-drain distance, justifying our exclusion of any temperature difference over the interface between the metallic contacts and the graphene flake.

\begin{figure}[h]
\begin{center}\leavevmode
\includegraphics[width=0.9\linewidth]{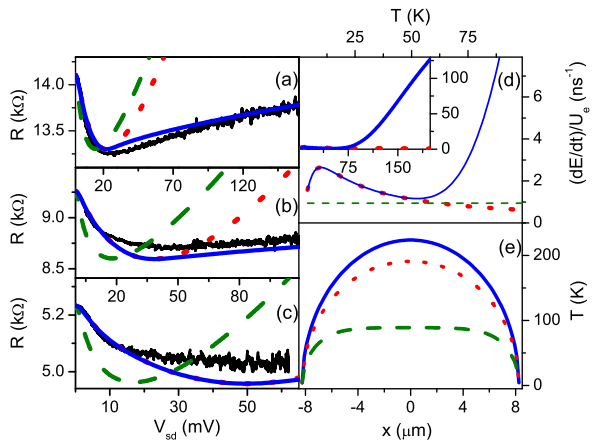}
\caption{(color online). (a)--(c) Measurements of $R(V_{sd})$ (black) in sample A for different gate voltages: $V_G$ = 3, -6, -12 V, respectively. The smooth curves are calculations for the case of no phonons (dashed green), acoustic phonons (dotted red), and both acoustic and optical phonons (blue). Note that in (c) the dotted and solid curves overlap. (d) The rate of energy loss of electrons due to scattering by acoustic phonons (dotted red line), and scattering by both acoustic and remote optical phonons (solid blue line). The dashed green line shows the inverse diffusion time. (e) The calculated temperature distribution along the length of the sample for the three different models (same color coding) with an applied source-drain bias of 70 mV. The bath temperature is 4.2 K, and $V_G$ = 3 V.}
\label{Fig4v0}
\end{center}
\end{figure}

To probe electron-phonon relaxation, we have studied the non-linear differential resistance of a longer sample, Fig.~\ref{Fig4v0}. The fit based on Eqs.~(\ref{EqRKTx}) and (\ref{EqRVsdTxIntegral}) (dashed line) is clearly inadequate, as it results in a dependence that is steeper than that actually observed. This means that Eq.~(\ref{EqRKTx}) overestimates the electron temperature. The disagreement can be accounted for if we include cooling via phonon emission. Thus, phonon cooling is indeed important for the 16 $\mu$m sample.  We use a heat balance equation:

\begin{equation}
\frac{\partial Q}{\partial x} = \frac{V^2}{\rho L^2} + \left( \frac{d E}{dt} \right)_{ph},
\label{EqHeatBalance}
\end{equation}

\noindent where $Q = \kappa(T) dT_e/dx$ is the heat flux, $\kappa$ is the electronic thermal conductivity, $V^2/\rho L^2$ is the Joule heat, $\rho$ is the electrical resistivity, and $(dE/dt)_{ph}$ is the energy dissipation rate due to phonons. The electronic thermal conductivity is given by the Wiedemann-Franz law, $\kappa = \pi^2 k_B^2 T_e/3e^2 \rho$. We assume that the phonon temperature is equal to the bath temperature, and impose the boundary conditions $T_e(x = \pm L/2) = T_B$, which assumes complete cooling at the leads. Given the small non-linearity seen in the $R(V_{sd})$ at $T_B$ = 107 K (Fig.~\ref{PaperFig1}), one might assume that up to $V_{sd} \sim$ 150 mV the temperature of electrons does not significantly exceed 100 K. Transport measurements \cite{ChenNatNano08} have shown that optical phonons, either in the SiO$_2$ substrate or in the graphene lattice, do not contribute significantly to transport below $\sim$200 K. Thus, we begin by considering only acoustic phonons in the graphene lattice. The energy dissipation due to acoustical phonons was found in \cite{BistritzerPRL09,TsePRB09}:

\begin{equation}
\label{diss-dE}
\left(\frac{dE}{dt} \right)_{ac}= - \frac{D_a^2 k_B E_F^5}{2 \varrho \hbar^5 v_F^8}T_{e},
\end{equation}

\noindent where $D_a$ = 18 -- 21 eV is the deformation potential, $\varrho = 7.6\times 10^{-7}\,\mathrm{kgm^{-2}}$ and $v_s = 2\times 10^4\,\mathrm{ms^{-1}}$ are the areal mass density and speed of sound in monolayer graphene \cite{HwangPRB08}. The ratio of this function to the electron energy density $U_e \propto T_e^2$ gives the characteristic rate of temperature equilibration of the electrons via phonon scattering, which is plotted in Fig.~\ref{Fig4v0}(d). The nonmonotonic behavior of the relaxation rate occurs due to cooling by acoustic phonons becoming ineffective at high temperatures: since the energy of acoustic phonons is limited to the Bloch-Gruneisen temperature, $k_B T_{BG}$, multiple scattering events are required to equilibrate hot electrons ($T_e > T_{BG}$). In Fig.~\ref{Fig4v0}(d), where $V_G$ = 3 V, this corresponds to $T_{BG}$ = 25 K, and we see that the maximum in $(dE/dt)_{ac}/U_e$ occurs at $T_e \sim T_{BG}/2$. Also shown in the figure is the inverse diffusion time, $\tau_D^{-1} = \pi^2 D/L^2$, which relates to the typical time for hot carriers to diffuse out from the sample center into the metallic contacts. In the 16$\mu$m sample this mechanism plays a minor role relative to the cooling by electron-phonon scattering at all temperatures.

The temperature profile in the sample can be determined through Eqs.~(\ref{EqHeatBalance}) and (\ref{diss-dE}). This is plotted in Fig.~\ref{Fig4v0}(e) for an applied bias of 70 mV. Using this $T(x)$ for each $V_{sd}$ in Eq.~(\ref{EqRVsdTxIntegral}), the resulting $R(V_{sd})$ are plotted in Fig.~\ref{Fig4v0}(a)--(c), with $V_G$ = 3, -6, and -12 V, respectively. We take $D_a = 18\, \rm eV$ for $V_G$ = 3 V, and $D_a = 21\, \rm eV$ for $V_G$ = -6 and -12 V \cite{DeanNatNanoTech10}. In comparison to purely electronic heat transfer, the low-bias region is well described by the inclusion of acoustic phonon scattering, correctly identifying the point in $V_{sd}$ where the up-turn in resistance occurs. However, at large biases, where cooling by acoustic phonons is inefficient, our model predicts a steep increase in $R(V_{sd})$, in striking contrast to the data. This indicates that an additional cooling mechanism becomes dominant at higher electron temperatures, $T>70$ K.

A plausible mechanism for carriers to dissipate heat is via the RIPs in the SiO$_2$ substrate, which have an unusually low frequency, $\hbar \omega_s$ = 59 meV. Scattering by these optical phonon modes has been proposed to explain the temperature dependence of the graphene resistance at $T>$200 K.\cite{ChenNatNano08} These polar optical phonons in the substrate couple with carriers in the sample via a fringing electric field, which decays exponentially away from the substrate surface. The scattering by these phonons can be significant in graphene samples due to the close proximity of the carriers in graphene to the underlying substrate. We have found the corresponding contribution to the cooling power \cite{SOM}:

\begin{equation}
\label{Edotopt}
\left(\frac{dE}{dt}\right)_{opt} \sim \frac{E_F e^2 \beta \omega_s^3}{\hbar v_F}\exp\left(-\frac{\hbar \omega_s}{k_B T}\right),
\end{equation}

\noindent
where $\beta = 0.025$ is a dimensionless coupling constant \cite{SOM,FratiniPRB08}, which is related to the dielectric function $\epsilon(\omega)$ of the substrate. Comparing this to Eq.~(\ref{diss-dE}), one can see that optical phonons dominate the energy dissipation, see Fig.~\ref{Fig4v0}(d). We have used Eq.~(\ref{EqHeatBalance}) with the two dissipation terms Eqs. (\ref{diss-dE}) and (\ref{Edotopt}) to determine the temperature distribution in the high-current regime. A steep exponential increase in dissipation stabilizes the electron temperature and this gives rise to a flat temperature profile, Fig.~\ref{Fig4v0}(e). The resulting $R(V_{sd})$ (solid blue lines in Fig.~\ref{Fig4v0}(a)--(c)) now give a reasonable fit to the measured values over the entire range of source drain bias.

To summarize, we have measured the differential resistance of monolayer graphene in response to an applied DC source-drain bias at low temperatures. We observe a peak in the resistance around zero bias, with an upturn at higher biases. We demonstrate that this structure arises from self-heating of charge carriers. By modeling the behavior of the resistance as a function of bias, we demonstrate that surface phonons of the underlying substrate are the dominant heat dissipation mechanism for carriers at carrier temperatures above $\sim$70 K.

We are grateful to F. Withers for providing some of the graphene samples, and P.R. Wilkins for technical assistance. This work was supported under EPSRC grant number EP/G041482/1. DWH and AS would also like to acknowledge support from the EPSRC grant EP/G036101/1.

\bibliography{LTJH}

\end{document}